\documentclass[%
 reprint,
 amsmath,amssymb,
 aps,
prl,
]{revtex4-2}

\usepackage{graphicx}
\usepackage{float}
\usepackage{verbatim}
\usepackage{dcolumn}
\usepackage{physics}
\usepackage{bm}
\usepackage{lipsum}
\usepackage{xcolor}
\usepackage{soul}
\setstcolor{red}

\begin{document}

\preprint{APS/123-QED}

\title{Sensing quantum vacuum fluctuations with non-Gaussian electronic noise}

\author{Clovis Farley}
\author{Edouard Pinsolle}
\author{Bertrand Reulet}%


\affiliation{Département de physique, Institut quantique, Université de Sherbrooke, Sherbrooke, Québec, Canada J1K 2R1
}%

\date{\today}

\begin{abstract}


The statistics of electron transport in a quantum conductor is affected by fluctuations of its voltage bias. Here we show experimentally how a third order correlation in the electromagnetic field arises from the noise of a tunnel junction in the microwave domain being modulated by the vacuum fluctuations generated by a resistor at ultralow temperature. This provides a way to measure the vacuum fluctuations experienced by the junction, not offset by the unavoidable noise added by the detection setup.

\end{abstract}

\maketitle

\emph{Introduction.}---Vacuum fluctuations (VF), the fact that the electromagnetic field always fluctuates, even in free space at zero temperature, is a direct consequence of the Heisenberg uncertainty principle, and as such a hallmark of quantum physics. Two consequences are found in most textbooks when solving the hydrogen atom: the energy levels are shifted (Lamb shift), even the ground state, and all excited states acquire a finite lifetime (spontaneous emission) (for a review on VF, see \cite{Milonni_book}). 

Mesoscopic circuits cooled to ultra-low temperatures and connected to microwave transmission lines bare many similarities with the hydrogen atom in vacuum, to the point that some simple circuits are nicknamed "artificial atoms"\cite{Blais2021}. Indeed, such circuits with a few degrees of freedom are well described by hamiltonians with a discrete spectrum, and a semi-infinite transmission line (or in practice, terminated by a matched load at low temperature) is nothing but a unidimensional vacuum for electromagnetic waves in the microwave domain\cite{Callen1951}. The energy levels of these circuits are shifted and have finite lifetime, a major difficulty for the development of solid state quantum computers \cite{Blais2021}.

While superconducting circuits such as those used as qubits can be considered similar to atoms, dissipative circuits made of normal conductors, by involving the electronic degrees of freedom, are much more complex, with continuous spectra. Among them the tunnel junction, made of two metallic contacts separated by a thin insulating film through which electrons can tunnel, is among the simplest. If purely voltage biased, it has a linear current-voltage characteristic like a simple resistor. Connected to a high impedance electromagnetic environment, it becomes nonlinear, a phenomenon called Dynamical Coulomb Blockade (DCB)\cite{Delsing1989, Ingold_Nazarov}, that in some sense is similar to spontaneous emission \cite{Altimiras2014}.

A resistor placed at very low temperature generates vacuum fluctuations into the coax cable connected to it. These can be amplified, however the amplifier itself adds its own noise to that of the resistor (and even often dominates the total signal), and there is no way to separate them. By replacing the resistor by a tunnel junction, which noise can be tuned by a dc bias voltage, from VF at zero bias to pure shot noise at high bias, the spectrum of VF has been measured up to 10 GHz \cite{Thibault15}. At much higher frequencies, a Josephson junction has been used to downconvert the high frequency ($\sim100$ GHz) VF of its shunt resistor to low frequency noise ($\sim100$ kHz) \cite{Koch81}.

Intrinsic to VF is the problem of their detection: one may want to distinguish processes where the electromagnetic field is detected by absorption or emission of energy. A system in its ground state cannot absorb energy from vacuum, while fluctuations may cause stimulated emission if the detector is in an excited state \cite{Intro_QNoise}. Different detection schemes correspond to different current-current correlators, since the quantum current operators at different times do not commute \cite{Gavish, Aguado}. Here we concentrate on detecting the field, not the energy associated to it. This can be performed by a detector driven out of equilibrium, which is sensitive to VF. The corresponding current-current correlator is the symmetrized one \cite{Qtape}. In optics it is commonly achieved by homodyne detection, where the bias is provided by a strong, classical electromagnetic field (see e.g. \cite{barnett_methods_2002}); in microwave by amplifiers using strongly dc biased field effect transistors \cite{Bozy, Eichler12a, Eichler12b}.

In this article we explore the effect of VF on the statistics of the current fluctuations generated by a tunnel junction beyond the average current and variance. The third moment of voltage fluctuations is known to be strongly influenced by the environment of the junction, both through its impedance and noise temperature \cite{Reulet2003,Beenakker2003,Gershon2008}. We show in this Letter that it is sensitive to VF, which modulate the noise of the junction.

\begin{figure}[h]
    \centering
    \includegraphics[width=8.6cm]{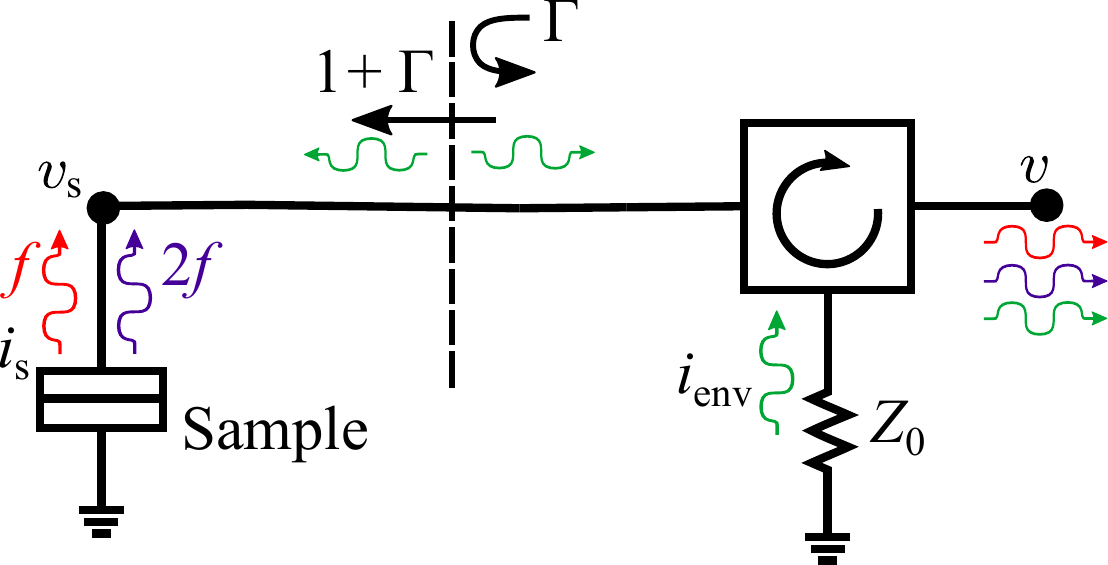}
    \caption{Principle of the detection scheme. Thanks to the circulator, the sample experiences the fluctuations generated by the load $Z_0$ (green arrows), which it partially reflects. These induce voltage fluctuations $v_\text{s}(t)$ across the sample, which modulate its noise. The detected voltage $v(t)$ contains the noise generated by the sample at frequencies $f$ (red) and $2f$ (blue) as well as the reflected environmental fluctuations at both frequencies (green). }
    \label{fig:simple}
\end{figure}

\emph{Theory.}---How the statistics of current fluctuations in a mesoscopic conductor translates into that of the detected voltage has been shown to be nontrivial  beyond the variance \cite{Reulet2003,Beenakker2003}. In order to give a clear physical picture of why, we start by omitting frequency dependence in the expressions. We consider the schematics depicted in Fig.\ref{fig:simple}. A fluctuating current $i_\text{s}(t)$ generated by the sample leads to a detected voltage $v(t)=\frac{1}{2}Z_0(1+\Gamma)i_\text{s}(t)$ with $Z_0$ the characteristic impedance of the transmission line between the sample and the amplifier matched to it (here $Z_0=50$ $\Omega$), and $\Gamma=(R_\text{s}-Z_0)/(R_\text{s}+Z_0)$ the voltage reflection coefficient of a wave impinging onto the sample of resistance $R_\text{s}$. 

The electromagnetic environment of the sample, modeled by a noise source of impedance $Z_0$ generating current fluctuations $i_{\text{env}}(t)$ also contributes to the detected voltage by being partially reflected by the sample, i.e. as $\frac{1}{2} Z_0\Gamma i_{\text{env}}$. As a result the third moment of the detected voltage contains, in average over time, two terms:
\begin{equation}
    \langle v^3\rangle=k\left[\langle i_{\text{s}}^3\rangle+ 3 \frac{\Gamma}{1+\Gamma} \langle i_{\text{s}}^2i_{\text{env}}\rangle \right]
\end{equation}
with $k=\frac{Z_0^3}{8}(1+\Gamma)^3$. The first term, $\langle i_{\text{s}}^3\rangle$ contains the intrinsic third moment of current fluctuations and the self modulation of the noise, also called feedback term, explained below. The second term corresponds to the modulation of the sample current fluctuations by the time-dependent voltage $v_{\text{s}}(t)$ across the sample coming from the noisy environment. It is given by $\chi\langle v_{\text{s}} i_{\text{env}}\rangle$,  as obtained by cascaded averaging \cite{Beenakker2003, Nagaev2002}. Here $\chi\sim d\langle i_{\text{s}}^2\rangle/d{V_{\text{s}}}$ is the noise susceptibility, which measures how much a small voltage variation modifies the variance of current fluctuations \cite{Gabelli2008,Gabelli_SPIE}. $V_{\text{s}}$ is the dc voltage bias on the junction and $\langle v_{\text{s}} i_{\text{env}}\rangle= \frac{1}{2} (1+\Gamma)Z_0\langle i_{\text{env}}^2\rangle$. 
This mechanism is at the heart of our detection scheme: $i_{\text{env}}$ modulates the noise generated by the sample, their correlations leading to a third moment. In that sense, the sample acts as an on-chip voltage-to-noise transducer.
The feedback term mentioned above corresponds to the same mechanism, replacing $i_{\text{env}}$ by $i_\text{s}$. While $\langle i_{\text{s}}^3\rangle$ and $\chi$ may have complex frequency dependence in the quantum regime at low voltage \cite{Gabelli2008, Gabelli_SPIE, Gabelli2013, Farley2023, Galaktionov2003, Salo2006, Bednorz2010}, they are simple in the classical limit of pure shot noise, when $e V_{\text{s}} \gg hf, k_B T$ with $T$ the sample temperature. In this limit the electrons follow a Poisson statistics with current fluctuations having spectral densities given by $e \abs{I_{\text{s}}} $ for the variance~\cite{Blanter2000} and $e^2I_{\text{s}}$ for the third moment, both frequency independent. $I_{\text{s}}=V_\text{s}/R_\text{s}$ is the average, dc bias current in the junction. The noise susceptibility is simply $\chi=d(e \abs{I_{\text{s}}} )/dV_s= \text{sign}(I_\text{s}) e/R_\text{s}$. Our experiment corresponds to the situation where the sample is classical but the electromagnetic environment is in the vacuum state.

We now introduce the frequency dependence. The spectral density of the third moment of the detected voltage fluctuations involves three frequencies, $\langle v(f_1) v(f_2) v(f_3)\rangle$. Time-averaging imposes $f_1+f_2+f_3=0$. In the following we will consider the special case where only two frequencies are involved in the detection, $f_1=f_2=f$, $f_3=-2f$, which simplifies the experiment. Restoring the frequency dependence of the quantities, we find for $\abs{eV_{\text{s}}} \gg hf, k_B T$:

\begin{equation}
    \langle v(f)^2v(-2f)\rangle = k \left[(3x-1)e^2I_{\text{s}} + \text{sign}(I_\text{s})\frac{\Gamma}{2} \frac{Z_0}{R_\text{s}}  e  \tilde{S}_{\text{env}}(f)\right]
    \label{eq:S3mes}
\end{equation}
with $x=\frac{1}{2}(Z_0/R_\text{s})(1+\Gamma)=Z_0/(R_\text{s}+Z_0)$ and
\begin{equation}
    \tilde{S}_{\text{env}}(f)=2S_{\text{eq}}(f)+S_{\text{eq}}(2f)
    \label{eq:Senv}
\end{equation}
with $S_{\text{eq}}(f)=(hf/Z_0)\coth{(hf/2k_BT_{\text{env}})}$ the current spectral density of equilibrium noise, thermal and quantum, of a resistor $Z_0$ at temperature $T_\textrm{env}$ \cite{Callen1951} (this derivation uses cascaded averaging \cite{Beenakker2003, Nagaev2002}, but a fully quantum treatment is still lacking). At high temperature ($k_B T_{\text{env}} \gg hf$), $\tilde{S}_{\text{env}}\simeq 6k_BT_{\text{env}}/Z_0$ corresponds to thermal noise while at low temperature, $ \tilde{S}_{\text{env}}\simeq 4hf/Z_0$ reduces to VF. As obvious in Eq. (\ref{eq:S3mes}), the third moment of the detected voltage at frequencies $(f,-2f)$ is a linear function of the bias current with an offset proportional to  $\tilde{S}_{\text{env}}$, i.e. $\ev{v(f)^2 v(-2f)}\propto Z_0^3e^2[I_\text{s}-\text{sign}(I_\text{s})I_0]$ with:
\begin{equation}
    eI_0= \frac{\Gamma Z_0}{2R_\text{s}(1-3x)} \tilde{S}_{\text{env}} .
    \label{eq:I0}
\end{equation}
The offset $I_0$ is proportional to the environmental noise experienced by the junction.  When the environment is in the quantum limit, $I_0$ is a measure of VF. The prefactor in Eq.(\ref{eq:I0}) involves the resistance and reflection coefficient of the sample, that can be determined separately. It does not involve parameters of the detection such as its gain.

\begin{figure}
    \centering
    \includegraphics[width=8.6cm]{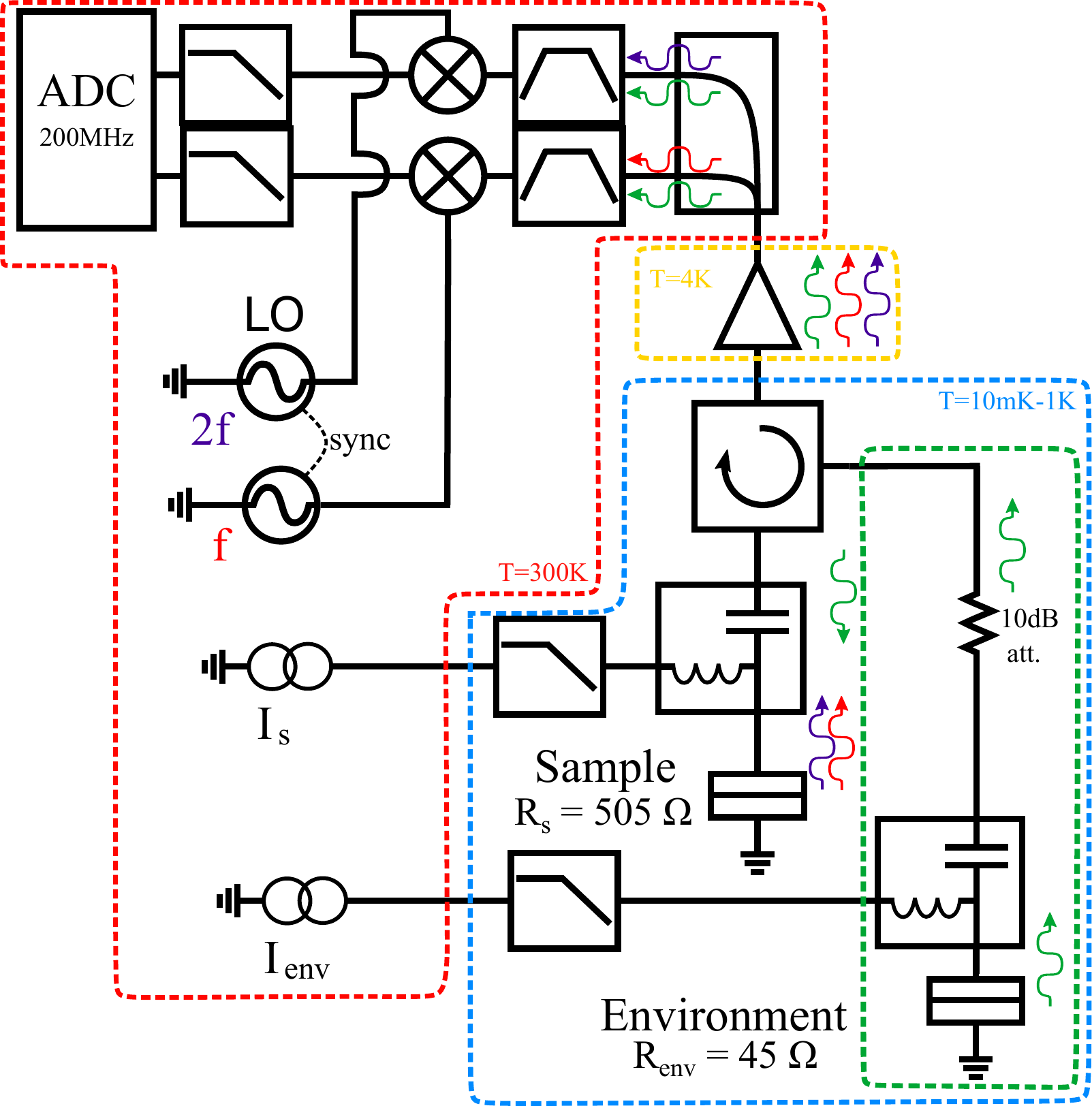}
    \caption{Experimental setup for experiment I. The dashed blue box indicates the low temperature stage of the dilution refrigerator, which temperature can be controlled from 10 mK to 1 K. The dashed green box is the electromagnetic environment, which noise influences the sample. In experiment II, it is replaced by a 50 $\Omega$ resistor with no connection to room temperature.}
    \label{fig:exp_setup}
\end{figure}

\emph{Experimental setup.}---It is usually best for microwave measurements to have the sample matched to the transmission lines, i.e. $\Gamma=0$. Unfortunately, this washes out the effect we seek to measure, see Eq.(\ref{eq:S3mes}): VF generated by the environment must be reflected by the sample to be amplified and detected. Our sample is a 2 \textmu m$\times$2 \textmu m Al-Al$_2$O$_3$ tunnel junction of dc resistance $R_\text{s}=505$~$\Omega$ i.e., $\Gamma = 0.82$ (DCB is negligible here given the low impedance, $50 $ $\Omega$ environment). The junction is fabricated on a high resistivity, oxidized Si substrate in a single photo-lithography step by shadow evaporation of Al through a 1 \textmu m high Dolan bridge \cite{Dolan1977} using e-beam deposition \cite{Spietz2003}. The junction is placed between the center conductor and ground plane of a 50 $\Omega$ CPW transmission line made of 300 nm thick Al, wire bonded to a microwave sample holder. A strong permanent magnet placed under the sample holder keeps the Al into its normal, non-superconducting state.

The setup is shown in Fig. \ref{fig:exp_setup}. The sample holder is anchored on the 10 mK stage of a dilution refrigerator. A bias-tee separates the dc current bias from the high frequency noise generated by the junction. A two-stage 4-12 GHz circulator provides 36 dB of isolation  between the 4 K plate of the refrigerator and the junction. It insures that, at the detection frequencies, the sample experiences only VF at the lowest temperatures, and not an uncontrolled noise, such as the one coming from the amplifier. The sample noise together with the environmental noise reflected by the sample are amplified through a 1-12~GHz HEMT amplifier placed at 4 K. At room temperature, this signal is split into two frequency bands around $f=5.05$ GHz and $2f=10.1$ GHz using a diplexer and filters, then downconverted using two mixers with phase-locked microwave sources as local oscillators at $f$ and $2f$, low-pass filtered below 117 MHz and digitized using synchronized 14 bits, 400 MSample/s analog-to-digital converters. In the following we will note $X_f(t)$ and $X_{2f}(t)$ the signals digitized after demodulation at frequencies $f$ and $2f$, i.e. the in-phase quadratures. 

After acquiring 256 MSample on both channels, the 2D histogram of $(X_f,X_{2f})$ is populated, from which we compute the relevant moments of the distribution of $X_f$ and $X_{2f}$ : the variances $\ev{X_f^2 }$ and $\ev{X_{2f}^2}$, and the correlator $\ev{X_{f}^2X_{2f} }$ which we will denote $\ev{X^3}$ for the sake of brevity. Note that $\expval{X_f^3}=\expval{X_{2f}^3}=0$: having a non-zero third moment of a quadrature at a single frequency requires an ac excitation of the sample \cite{Bozy, Eichler12a, Eichler12b,S3cyclo}. This procedure is repeated $\sim$ 5000 times for every value of $I_s$. The obtained moments are then averaged together to provide the final result while the error bars are computed from the standard deviation on the distribution of the measured moments. The absolute phase of the measurements has no influence since the sample does not experience any ac excitation, so $\ev{X_f^2 }=\ev{Y_f^2 }\propto S_2(f)$ with $Y$ the other quadrature of the signal and $S_2$ the noise spectral density. Similarly, $\ev{X_{2f}^2}=\ev{Y_{2f}^2}\propto S_2(2f)$. In contrast, the relative phase between the oscillators at $f$ and $2f$ matters. It is controlled by a programmable delay line. We chose it so that $\ev{X_{f}^2Y_{2f}} =0$, i.e. $\ev{X^3}$ is proportional to the real part of $\ev{v(f)^2v(-2f)}$ (for more details, see supp. mat.). While non-linearities in the setup have negligible effects on the measurement of $\ev{X^2}$, they add contributions to $ \ev{X^3} $, in particlar terms of order 4 such as $\ev{ X_f^2X_{2f}^2 }$. These terms are even functions of the bias current $I_{s}$. To remove them from $ \ev{X^3} $, which is an odd function of the bias current, we simply anti-symmetrize it with respect to $I_{\text{s}}$ \cite{Reulet2003, Gabelli2013}. One of the most limiting factor of our experiment is the noise of the HEMT amplifier, which heavily degrades our signal-to-noise ratio, thus considerably increasing the averaging time. A quantum-limited amplifier could potentially speed up the measurement of $\ev{X^3}$, however with potentials pitfalls coming from the strong non-linearities and frequency-dependent phase response of such amplifiers\cite{TWPAs}.

We have performed two experiments corresponding to different environments. In experiment I, the environment consists of a 10 dB attenuator followed by another tunnel junction, the \emph{environmental junction} (EJ, of resistance $R_{\text{env}} = 45\Omega)$, see Fig.\ref{fig:exp_setup}. In the absence of bias, this environment is equivalent to a resistor as in Eq.(\ref{eq:Senv}). A dc current $I_{\text{env}}$ applied to the EJ allows to tune its noise  while keeping the fridge at 10 mK. The 10 dB attenuator inserted between the junction and the circulator insures a good impedance match of the junction. As a result, the total noise $\tilde S_{\text{env}}$ experienced by the sample consists of the equilibrium noise of the attenuator plus the attenuated shot noise of the EJ. In the limit  $e V_{\text{env}} \gg hf$, this means that Eq.(\ref{eq:Senv}) has to be replaced by:
\begin{equation}
    \tilde S_{\text{env}}(f) = \alpha eI_\text{env}+(1-\alpha) [ 2S_{\text{eq}}(f) + S_{\text{eq}}(2f) ]
    \label{S_env_jct}
\end{equation}
with $\alpha$ the attenuation factor between the two junctions.  In experiment II,  the environment simply consists of a cryogenic, macroscopic $50$ $\Omega$ resistor thermalized on the base plate of the refrigerator. Its temperature is close to that of the fridge, which is controlled from 10~mK to 0.6~K. In this second experiment, the noise experienced by the sample is varied by changing the temperature of the refrigerator, thus also that of the sample.\\

\begin{figure}[h]
    \centering
    \includegraphics[]{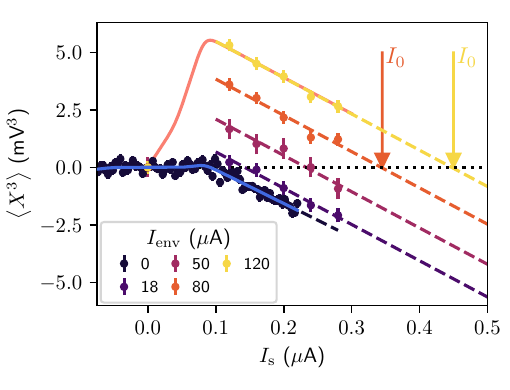}
    \caption{Examples of anti-symmetrized $\ev{X^3}$ vs. sample bias current $I_s$ for different environmental current biases $I_\text{env}$ in experiment I. Symbols are data, dashed lines are linear fits from which we extract $I_0$. The vertical arrows illustrate how $I_0$ increases with $I_{\text{env}}$. The solid lines are the theoretical prediction for $\ev{X^3}$ when the junction is exposed to VF ($I_{\text{env}}=0$, blue) or when $I_{\text{env}}$=120 $\mu$A (salmon).}
    \label{fig:varI}
\end{figure}

\emph{Results.}---Measurements of $ \ev{ X_f^2 } $ and $ \ev{ X_{2f}^2 } $ vs. sample bias current $I_s$ allow to determine the electron temperature $T$ of the sample \cite{Spietz2003, Spietz2006}. We find $T$ to be identical to the fridge temperature except at the lowest temperatures (we obtain $T=20$ mK at the base temperature, see supp. mat.). This is however irrelevant since we operate at high bias, where the sample is generating pure shot noise, i.e. $S_2(f)=S_2(2f)=e \abs{I_{\text{s}}}$. Measurements of $\ev{X^3}$ vs. $I_\text{s}$ in experiment I are reported in Fig. \ref{fig:varI}, for various biases of the EJ. For $I_\mathrm{env}=0$, we show the measurement of the full bias dependence, in very good agreement with theory (see supp. mat.). Here the only unknown parameter is the overall gain of the setup. In the following we consider only the classical, high bias limit of this curve, given by Eq.(\ref{eq:S3mes}). We observe that the high bias behaviour of $\ev{X^3}$ is linear with an offset, i.e. $\ev{X^3}\propto(I_\text{s}-I_0)$, as expected.

Acquiring the data shown in Fig.\ref{fig:varI} for $I_\text{env}=0$, i.e. with the full dependence on $I_\text{s}$ takes $\sim2$ weeks. While we considered important to do this measurement at the lowest temperature to make sure that the experiment is well controlled, it is irrelevant to the final goal of the experiment. To determine how $I_0$ depends on environmental noise, all is needed are a few points in the high $I_\text{s}$ regime. We show in Fig.\ref{fig:varI} a few examples of such a determination for various environmental bias currents (experiment I). As expected from Eq.(\ref{eq:S3mes}), changing $I_{\text{env}}$ only shifts the curves without changing their slope. We have repeated this procedure to determine $I_0$ as a function of $I_\text{env}$ (experiment I) or $T$ (experiments I and II).

\begin{figure}[h]
    \centering
    \includegraphics[width=8.6cm]{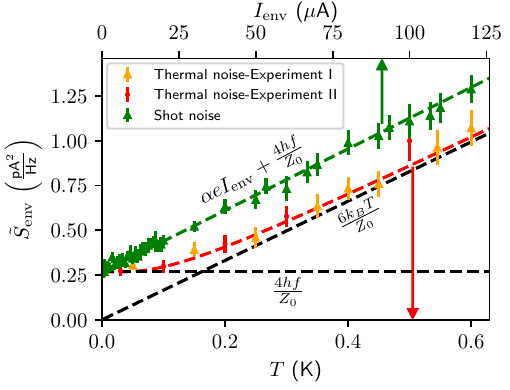}
    \caption{Environmental noise $\tilde{S}_{\text{env}}$ vs. environmental bias current $I_\text{env}$ (upper x scale) or temperature (lower x scale). Green triangles are measured by varying $I_{\text{env}}$ in experiment I, orange triangles by varying $T$ while keeping $I_{\text{env}}=0$ in experiment I and red dots by varying $T$ in experiment II, where the environment is a 50 $\Omega$ resistor.  The black horizontal dashed line corresponds to the theoretical limit of quantum VF and the oblique one to classical thermal noise. The red dotted line is the full theoretical prediction for $\tilde{S}_{\text{env}}$ of Eq.(\ref{eq:Senv}).}
    \label{fig:zpf}
\end{figure}

From the measurement of $I_0$ we deduce the environmental noise spectral density $\tilde{S}_{\text{env}}$ using Eq.(\ref{eq:I0}). Fig. \ref{fig:zpf} shows the measured dependence of $\tilde{S}_{\text{env}}$ with $I_{\text{env}}$ (between 0 and $120$ $\mu$A, experiment I) and the electron temperature $T$ (between 20 mK and 600 mK, experiments I and II).

We first discuss the dependence on $I_\textrm{env}$. The EJ is always biased in its classical limit so it generates pure shot noise $e\abs{I_\textrm{env}}$ at frequencies $f$ and $2f$ ($eV_\textrm{env}=hf$ corresponds to $I_\textrm{env}\simeq0.5$ $\mu$A). This noise is partially coupled to the microwave circuitry, attenuated by the 10 dB attenuator and further attenuated by the circuitry between the two junctions. As already discussed, $\tilde S_{\text{env}}$ grows linearly with $I_{\text{env}}$, according to Eq. (\ref{S_env_jct}), in good agreement with our observations. The observed slope $\alpha=0.0536$ corresponds to an effective attenuation of 17.5 dB between the EJ and the sample. An independent determination of $\alpha$ from $\ev{X_f^2}$ and $\ev{X_{2f}^2}$ measurements vs. $I_{\text{env}}$ at $I_\text{s}=0$ gives an  attenuation of 17.2 dB (see supp. mat.).

We now consider the same setup in which we do not bias the EJ but vary the temperature of the refrigerator. As previously, the sample is always biased in its classical regime $e\abs{V_s}\gg k_B T$, so $I_\text{s}$ is increased up to $I_\text{s} = 2.6$~$\mu$A for $T = 600$ mK. The corresponding measured $\tilde S_{\text{env}}$ are shown in Fig.\ref{fig:zpf}: it goes from a linear temperature dependence with no offset at high $T$  to a saturation at low $T$. This corresponds to the environmental noise going from thermal at high $T$ to VF at low $T$. The theoretical prediction of Eq.(\ref{eq:Senv}), shown with no fitting parameter in Fig. \ref{fig:zpf}, is in very good agreement with the measurement.

From the bias-dependence of the noise of the junctions \cite{Spietz2003,Spietz2006}, we find at base temperature $T=20$ mK for the sample and $T=23$ mK for the environmnetal junction, see sup. mat.. In experiment II, the environmental noise source is a macroscopic resistor well thermalized to the refrigerator. These data corroborate what has been measured in experiment I, see Fig.\ref{fig:zpf}.   

One may wonder whether one could modulate the VF to help their detection, as a signal turned on and off \cite{Bozy, Eichler12a, Eichler12b}, by e.g. connecting / disconnecting a resistor at the input of an amplifier. A matched resistor ($\Gamma=0$ in Fig. \ref{fig:simple}) leads to $\ev{v^2}\propto\ev{i_s^2}=S_{eq}(T)$ while for an open circuit ($\Gamma=1$) $\ev{v^2}\propto\ev{i_{env}^2}=S_{eq}(T_\text{env})$: if the measurement frequency is much larger than $k_BT/h$ and $k_BT_\text{env}/h$, in both cases one measures VF, that of the resistor or that of the load of the circulator, which are equal. This generalizes easily to any impedances.

\emph{Conclusion.}---We have measured the impact of a controlled electromagnetic noise on the current statistics of a tunnel junction by studying its effect on the third moment. Using a detection scheme providing an in-situ measurement of noise at cryogenic temperatures, we have observed that even the unavoidable quantum VF have a strong influence on electron transport beyond the usual dc current and variance. Indeed, their contribution to the skewness is almost as large as the intrinsic poissonian value  when $eV_s=2hf$ (see supp. mat.). This opens up the path to study the impact of quantum states of the electromagnetic field (vacuum, squeezed state, etc.) \cite{Souquet2014} on electron transport in a wide variety of devices and regimes.\\

\emph{Acknowledgments.}---We thank Pierre Février, Wolfgang Belzig, Julien Gabelli and Jérôme Estève for fruitful discussions. We are very grateful to Gabriel Laliberté and Christian Lupien for their technical help. This work was supported by the Canada Research Chair program, the NSERC, the Canada First Research Excellence Fund, the FRQNT, and the Canada Foundation for Innovation.

\bibliography{biblio}

\providecommand{\noopsort}[1]{}\providecommand{\singleletter}[1]{#1}%
\begin{thebibliography}{34}%
\makeatletter
\providecommand \@ifxundefined [1]{%
 \@ifx{#1\undefined}
}%
\providecommand \@ifnum [1]{%
 \ifnum #1\expandafter \@firstoftwo
 \else \expandafter \@secondoftwo
 \fi
}%
\providecommand \@ifx [1]{%
 \ifx #1\expandafter \@firstoftwo
 \else \expandafter \@secondoftwo
 \fi
}%
\providecommand \natexlab [1]{#1}%
\providecommand \enquote  [1]{``#1''}%
\providecommand \bibnamefont  [1]{#1}%
\providecommand \bibfnamefont [1]{#1}%
\providecommand \citenamefont [1]{#1}%
\providecommand \href@noop [0]{\@secondoftwo}%
\providecommand \href [0]{\begingroup \@sanitize@url \@href}%
\providecommand \@href[1]{\@@startlink{#1}\@@href}%
\providecommand \@@href[1]{\endgroup#1\@@endlink}%
\providecommand \@sanitize@url [0]{\catcode `\\12\catcode `\$12\catcode
  `\&12\catcode `\#12\catcode `\^12\catcode `\_12\catcode `\%12\relax}%
\providecommand \@@startlink[1]{}%
\providecommand \@@endlink[0]{}%
\providecommand \url  [0]{\begingroup\@sanitize@url \@url }%
\providecommand \@url [1]{\endgroup\@href {#1}{\urlprefix }}%
\providecommand \urlprefix  [0]{URL }%
\providecommand \Eprint [0]{\href }%
\providecommand \doibase [0]{https://doi.org/}%
\providecommand \selectlanguage [0]{\@gobble}%
\providecommand \bibinfo  [0]{\@secondoftwo}%
\providecommand \bibfield  [0]{\@secondoftwo}%
\providecommand \translation [1]{[#1]}%
\providecommand \BibitemOpen [0]{}%
\providecommand \bibitemStop [0]{}%
\providecommand \bibitemNoStop [0]{.\EOS\space}%
\providecommand \EOS [0]{\spacefactor3000\relax}%
\providecommand \BibitemShut  [1]{\csname bibitem#1\endcsname}%
\let\auto@bib@innerbib\@empty
\bibitem [{\citenamefont {Milonni}(1993)}]{Milonni_book}%
  \BibitemOpen
  \bibfield  {author} {\bibinfo {author} {\bibfnamefont {P.}~\bibnamefont
  {Milonni}},\ }\href@noop {} {\emph {\bibinfo {title} {The Quantum Vacuum}}}\
  (\bibinfo  {publisher} {Academic Press},\ \bibinfo {year} {1993})\BibitemShut
  {NoStop}%
\bibitem [{\citenamefont {Blais}\ \emph {et~al.}(2021)\citenamefont {Blais},
  \citenamefont {Grimsmo}, \citenamefont {Girvin},\ and\ \citenamefont
  {Wallraff}}]{Blais2021}%
  \BibitemOpen
  \bibfield  {author} {\bibinfo {author} {\bibfnamefont {A.}~\bibnamefont
  {Blais}}, \bibinfo {author} {\bibfnamefont {A.~L.}\ \bibnamefont {Grimsmo}},
  \bibinfo {author} {\bibfnamefont {S.~M.}\ \bibnamefont {Girvin}},\ and\
  \bibinfo {author} {\bibfnamefont {A.}~\bibnamefont {Wallraff}},\ }\bibfield
  {title} {\bibinfo {title} {Circuit quantum electrodynamics},\ }\href
  {https://doi.org/10.1103/RevModPhys.93.025005} {\bibfield  {journal}
  {\bibinfo  {journal} {Rev. Mod. Phys.}\ }\textbf {\bibinfo {volume} {93}},\
  \bibinfo {pages} {025005} (\bibinfo {year} {2021})}\BibitemShut {NoStop}%
\bibitem [{\citenamefont {Callen}\ and\ \citenamefont
  {Welton}(1951)}]{Callen1951}%
  \BibitemOpen
  \bibfield  {author} {\bibinfo {author} {\bibfnamefont {H.~B.}\ \bibnamefont
  {Callen}}\ and\ \bibinfo {author} {\bibfnamefont {T.~A.}\ \bibnamefont
  {Welton}},\ }\bibfield  {title} {\bibinfo {title} {Irreversibility and
  generalized noise},\ }\href {https://doi.org/10.1103/PhysRev.83.34}
  {\bibfield  {journal} {\bibinfo  {journal} {Phys. Rev.}\ }\textbf {\bibinfo
  {volume} {83}},\ \bibinfo {pages} {34} (\bibinfo {year} {1951})}\BibitemShut
  {NoStop}%
\bibitem [{\citenamefont {Delsing}\ \emph {et~al.}(1989)\citenamefont
  {Delsing}, \citenamefont {Likharev}, \citenamefont {Kuzmin},\ and\
  \citenamefont {Claeson}}]{Delsing1989}%
  \BibitemOpen
  \bibfield  {author} {\bibinfo {author} {\bibfnamefont {P.}~\bibnamefont
  {Delsing}}, \bibinfo {author} {\bibfnamefont {K.~K.}\ \bibnamefont
  {Likharev}}, \bibinfo {author} {\bibfnamefont {L.~S.}\ \bibnamefont
  {Kuzmin}},\ and\ \bibinfo {author} {\bibfnamefont {T.}~\bibnamefont
  {Claeson}},\ }\bibfield  {title} {\bibinfo {title} {Effect of high-frequency
  electrodynamic environment on the single-electron tunneling in ultrasmall
  junctions},\ }\href {https://doi.org/10.1103/PhysRevLett.63.1180} {\bibfield
  {journal} {\bibinfo  {journal} {Phys. Rev. Lett.}\ }\textbf {\bibinfo
  {volume} {63}},\ \bibinfo {pages} {1180} (\bibinfo {year}
  {1989})}\BibitemShut {NoStop}%
\bibitem [{\citenamefont {Ingold}\ and\ \citenamefont
  {Nazarov}(1992)}]{Ingold_Nazarov}%
  \BibitemOpen
  \bibfield  {author} {\bibinfo {author} {\bibfnamefont {G.-L.}\ \bibnamefont
  {Ingold}}\ and\ \bibinfo {author} {\bibfnamefont {Y.~V.}\ \bibnamefont
  {Nazarov}},\ }\bibinfo {title} {Charge tunneling rates in ultrasmall
  junctions},\ in\ \href {https://doi.org/10.1007/978-1-4757-2166-9_2} {\emph
  {\bibinfo {booktitle} {Single Charge Tunneling: Coulomb Blockade Phenomena In
  Nanostructures}}},\ \bibinfo {editor} {edited by\ \bibinfo {editor}
  {\bibfnamefont {H.}~\bibnamefont {Grabert}}\ and\ \bibinfo {editor}
  {\bibfnamefont {M.~H.}\ \bibnamefont {Devoret}}}\ (\bibinfo  {publisher}
  {Springer US},\ \bibinfo {address} {Boston, MA},\ \bibinfo {year} {1992})\
  pp.\ \bibinfo {pages} {21--107}\BibitemShut {NoStop}%
\bibitem [{\citenamefont {Altimiras}\ \emph {et~al.}(2014)\citenamefont
  {Altimiras}, \citenamefont {Parlavecchio}, \citenamefont {Joyez},
  \citenamefont {Vion}, \citenamefont {Roche}, \citenamefont {Esteve},\ and\
  \citenamefont {Portier}}]{Altimiras2014}%
  \BibitemOpen
  \bibfield  {author} {\bibinfo {author} {\bibfnamefont {C.}~\bibnamefont
  {Altimiras}}, \bibinfo {author} {\bibfnamefont {O.}~\bibnamefont
  {Parlavecchio}}, \bibinfo {author} {\bibfnamefont {P.}~\bibnamefont {Joyez}},
  \bibinfo {author} {\bibfnamefont {D.}~\bibnamefont {Vion}}, \bibinfo {author}
  {\bibfnamefont {P.}~\bibnamefont {Roche}}, \bibinfo {author} {\bibfnamefont
  {D.}~\bibnamefont {Esteve}},\ and\ \bibinfo {author} {\bibfnamefont
  {F.}~\bibnamefont {Portier}},\ }\bibfield  {title} {\bibinfo {title}
  {Dynamical coulomb blockade of shot noise},\ }\href
  {https://doi.org/10.1103/PhysRevLett.112.236803} {\bibfield  {journal}
  {\bibinfo  {journal} {Phys. Rev. Lett.}\ }\textbf {\bibinfo {volume} {112}},\
  \bibinfo {pages} {236803} (\bibinfo {year} {2014})}\BibitemShut {NoStop}%
\bibitem [{\citenamefont {Thibault}\ \emph {et~al.}(2015)\citenamefont
  {Thibault}, \citenamefont {Gabelli}, \citenamefont {Lupien},\ and\
  \citenamefont {Reulet}}]{Thibault15}%
  \BibitemOpen
  \bibfield  {author} {\bibinfo {author} {\bibfnamefont {K.}~\bibnamefont
  {Thibault}}, \bibinfo {author} {\bibfnamefont {J.}~\bibnamefont {Gabelli}},
  \bibinfo {author} {\bibfnamefont {C.}~\bibnamefont {Lupien}},\ and\ \bibinfo
  {author} {\bibfnamefont {B.}~\bibnamefont {Reulet}},\ }\bibfield  {title}
  {\bibinfo {title} {Pauli-heisenberg oscillations in electron quantum
  transport},\ }\href {https://doi.org/10.1103/PhysRevLett.114.236604}
  {\bibfield  {journal} {\bibinfo  {journal} {Phys. Rev. Lett.}\ }\textbf
  {\bibinfo {volume} {114}},\ \bibinfo {pages} {236604} (\bibinfo {year}
  {2015})}\BibitemShut {NoStop}%
\bibitem [{\citenamefont {Koch}\ \emph {et~al.}(1981)\citenamefont {Koch},
  \citenamefont {Van~Harlingen},\ and\ \citenamefont {Clarke}}]{Koch81}%
  \BibitemOpen
  \bibfield  {author} {\bibinfo {author} {\bibfnamefont {R.~H.}\ \bibnamefont
  {Koch}}, \bibinfo {author} {\bibfnamefont {D.~J.}\ \bibnamefont
  {Van~Harlingen}},\ and\ \bibinfo {author} {\bibfnamefont {J.}~\bibnamefont
  {Clarke}},\ }\bibfield  {title} {\bibinfo {title} {Observation of zero-point
  fluctuations in a resistively shunted josephson tunnel junction},\ }\href
  {https://doi.org/10.1103/PhysRevLett.47.1216} {\bibfield  {journal} {\bibinfo
   {journal} {Phys. Rev. Lett.}\ }\textbf {\bibinfo {volume} {47}},\ \bibinfo
  {pages} {1216} (\bibinfo {year} {1981})}\BibitemShut {NoStop}%
\bibitem [{\citenamefont {Clerk}\ \emph {et~al.}(2010)\citenamefont {Clerk},
  \citenamefont {Devoret}, \citenamefont {Girvin}, \citenamefont {Marquardt},\
  and\ \citenamefont {Schoelkopf}}]{Intro_QNoise}%
  \BibitemOpen
  \bibfield  {author} {\bibinfo {author} {\bibfnamefont {A.~A.}\ \bibnamefont
  {Clerk}}, \bibinfo {author} {\bibfnamefont {M.~H.}\ \bibnamefont {Devoret}},
  \bibinfo {author} {\bibfnamefont {S.~M.}\ \bibnamefont {Girvin}}, \bibinfo
  {author} {\bibfnamefont {F.}~\bibnamefont {Marquardt}},\ and\ \bibinfo
  {author} {\bibfnamefont {R.~J.}\ \bibnamefont {Schoelkopf}},\ }\bibfield
  {title} {\bibinfo {title} {Introduction to quantum noise, measurement, and
  amplification},\ }\href {https://doi.org/10.1103/RevModPhys.82.1155}
  {\bibfield  {journal} {\bibinfo  {journal} {Rev. Mod. Phys.}\ }\textbf
  {\bibinfo {volume} {82}},\ \bibinfo {pages} {1155} (\bibinfo {year}
  {2010})}\BibitemShut {NoStop}%
\bibitem [{\citenamefont {Gavish}\ \emph {et~al.}(2000)\citenamefont {Gavish},
  \citenamefont {Levinson},\ and\ \citenamefont {Imry}}]{Gavish}%
  \BibitemOpen
  \bibfield  {author} {\bibinfo {author} {\bibfnamefont {U.}~\bibnamefont
  {Gavish}}, \bibinfo {author} {\bibfnamefont {Y.}~\bibnamefont {Levinson}},\
  and\ \bibinfo {author} {\bibfnamefont {Y.}~\bibnamefont {Imry}},\ }\bibfield
  {title} {\bibinfo {title} {Detection of quantum noise},\ }\href
  {https://doi.org/10.1103/PhysRevB.62.R10637} {\bibfield  {journal} {\bibinfo
  {journal} {Phys. Rev. B}\ }\textbf {\bibinfo {volume} {62}},\ \bibinfo
  {pages} {R10637} (\bibinfo {year} {2000})}\BibitemShut {NoStop}%
\bibitem [{\citenamefont {Aguado}\ and\ \citenamefont
  {Kouwenhoven}(2000)}]{Aguado}%
  \BibitemOpen
  \bibfield  {author} {\bibinfo {author} {\bibfnamefont {R.}~\bibnamefont
  {Aguado}}\ and\ \bibinfo {author} {\bibfnamefont {L.~P.}\ \bibnamefont
  {Kouwenhoven}},\ }\bibfield  {title} {\bibinfo {title} {Double quantum dots
  as detectors of high-frequency quantum noise in mesoscopic conductors},\
  }\href {https://doi.org/10.1103/PhysRevLett.84.1986} {\bibfield  {journal}
  {\bibinfo  {journal} {Phys. Rev. Lett.}\ }\textbf {\bibinfo {volume} {84}},\
  \bibinfo {pages} {1986} (\bibinfo {year} {2000})}\BibitemShut {NoStop}%
\bibitem [{\citenamefont {Bednorz}\ and\ \citenamefont
  {Belzig}(2010{\natexlab{a}})}]{Qtape}%
  \BibitemOpen
  \bibfield  {author} {\bibinfo {author} {\bibfnamefont {A.}~\bibnamefont
  {Bednorz}}\ and\ \bibinfo {author} {\bibfnamefont {W.}~\bibnamefont
  {Belzig}},\ }\bibfield  {title} {\bibinfo {title} {Models of mesoscopic
  time-resolved current detection},\ }\href
  {https://doi.org/10.1103/PhysRevB.81.125112} {\bibfield  {journal} {\bibinfo
  {journal} {Phys. Rev. B}\ }\textbf {\bibinfo {volume} {81}},\ \bibinfo
  {pages} {125112} (\bibinfo {year} {2010}{\natexlab{a}})}\BibitemShut
  {NoStop}%
\bibitem [{\citenamefont {Barnett}\ and\ \citenamefont
  {Radmore}(2002)}]{barnett_methods_2002}%
  \BibitemOpen
  \bibfield  {author} {\bibinfo {author} {\bibfnamefont {S.}~\bibnamefont
  {Barnett}}\ and\ \bibinfo {author} {\bibfnamefont {P.}~\bibnamefont
  {Radmore}},\ }\href
  {https://doi.org/10.1093/acprof:oso/9780198563617.001.0001} {\emph {\bibinfo
  {title} {Methods in Theoretical Quantum Optics}}}\ (\bibinfo  {publisher}
  {Oxford University Press},\ \bibinfo {year} {2002})\BibitemShut {NoStop}%
\bibitem [{\citenamefont {Bozyigit}\ \emph {et~al.}(2011)\citenamefont
  {Bozyigit}, \citenamefont {Lang}, \citenamefont {Steffen}, \citenamefont
  {Fink}, \citenamefont {Eichler}, \citenamefont {Baur}, \citenamefont
  {Bianchetti}, \citenamefont {Leek}, \citenamefont {Filipp}, \citenamefont
  {da~Silva}, \citenamefont {Blais},\ and\ \citenamefont {Wallraff}}]{Bozy}%
  \BibitemOpen
  \bibfield  {author} {\bibinfo {author} {\bibfnamefont {D.}~\bibnamefont
  {Bozyigit}}, \bibinfo {author} {\bibfnamefont {C.}~\bibnamefont {Lang}},
  \bibinfo {author} {\bibfnamefont {L.}~\bibnamefont {Steffen}}, \bibinfo
  {author} {\bibfnamefont {J.}~\bibnamefont {Fink}}, \bibinfo {author}
  {\bibfnamefont {C.}~\bibnamefont {Eichler}}, \bibinfo {author} {\bibfnamefont
  {M.}~\bibnamefont {Baur}}, \bibinfo {author} {\bibfnamefont {R.}~\bibnamefont
  {Bianchetti}}, \bibinfo {author} {\bibfnamefont {P.}~\bibnamefont {Leek}},
  \bibinfo {author} {\bibfnamefont {S.}~\bibnamefont {Filipp}}, \bibinfo
  {author} {\bibfnamefont {M.}~\bibnamefont {da~Silva}}, \bibinfo {author}
  {\bibfnamefont {A.}~\bibnamefont {Blais}},\ and\ \bibinfo {author}
  {\bibfnamefont {A.}~\bibnamefont {Wallraff}},\ }\bibfield  {title} {\bibinfo
  {title} {Antibunching of microwave-frequency photons observed in correlation
  measurements using linear detectors},\ }\href@noop {} {\bibfield  {journal}
  {\bibinfo  {journal} {Nature Physics}\ }\textbf {\bibinfo {volume} {7}},\
  \bibinfo {pages} {154} (\bibinfo {year} {2011})}\BibitemShut {NoStop}%
\bibitem [{\citenamefont {Eichler}\ \emph
  {et~al.}(2012{\natexlab{a}})\citenamefont {Eichler}, \citenamefont
  {Bozyigit},\ and\ \citenamefont {Wallraff}}]{Eichler12a}%
  \BibitemOpen
  \bibfield  {author} {\bibinfo {author} {\bibfnamefont {C.}~\bibnamefont
  {Eichler}}, \bibinfo {author} {\bibfnamefont {D.}~\bibnamefont {Bozyigit}},\
  and\ \bibinfo {author} {\bibfnamefont {A.}~\bibnamefont {Wallraff}},\
  }\bibfield  {title} {\bibinfo {title} {Characterizing quantum microwave
  radiation and its entanglement with superconducting qubits using linear
  detectors},\ }\href {https://doi.org/10.1103/PhysRevA.86.032106} {\bibfield
  {journal} {\bibinfo  {journal} {Phys. Rev. A}\ }\textbf {\bibinfo {volume}
  {86}},\ \bibinfo {pages} {032106} (\bibinfo {year}
  {2012}{\natexlab{a}})}\BibitemShut {NoStop}%
\bibitem [{\citenamefont {Eichler}\ \emph
  {et~al.}(2012{\natexlab{b}})\citenamefont {Eichler}, \citenamefont {Lang},
  \citenamefont {Fink}, \citenamefont {Govenius}, \citenamefont {Filipp},\ and\
  \citenamefont {Wallraff}}]{Eichler12b}%
  \BibitemOpen
  \bibfield  {author} {\bibinfo {author} {\bibfnamefont {C.}~\bibnamefont
  {Eichler}}, \bibinfo {author} {\bibfnamefont {C.}~\bibnamefont {Lang}},
  \bibinfo {author} {\bibfnamefont {J.~M.}\ \bibnamefont {Fink}}, \bibinfo
  {author} {\bibfnamefont {J.}~\bibnamefont {Govenius}}, \bibinfo {author}
  {\bibfnamefont {S.}~\bibnamefont {Filipp}},\ and\ \bibinfo {author}
  {\bibfnamefont {A.}~\bibnamefont {Wallraff}},\ }\bibfield  {title} {\bibinfo
  {title} {Observation of entanglement between itinerant microwave photons and
  a superconducting qubit},\ }\href
  {https://doi.org/10.1103/PhysRevLett.109.240501} {\bibfield  {journal}
  {\bibinfo  {journal} {Phys. Rev. Lett.}\ }\textbf {\bibinfo {volume} {109}},\
  \bibinfo {pages} {240501} (\bibinfo {year} {2012}{\natexlab{b}})}\BibitemShut
  {NoStop}%
\bibitem [{\citenamefont {Reulet}\ \emph {et~al.}(2003)\citenamefont {Reulet},
  \citenamefont {Senzier},\ and\ \citenamefont {Prober}}]{Reulet2003}%
  \BibitemOpen
  \bibfield  {author} {\bibinfo {author} {\bibfnamefont {B.}~\bibnamefont
  {Reulet}}, \bibinfo {author} {\bibfnamefont {J.}~\bibnamefont {Senzier}},\
  and\ \bibinfo {author} {\bibfnamefont {D.~E.}\ \bibnamefont {Prober}},\
  }\bibfield  {title} {\bibinfo {title} {Environmental effects in the third
  moment of voltage fluctuations in a tunnel junction},\ }\href
  {https://doi.org/10.1103/PhysRevLett.91.196601} {\bibfield  {journal}
  {\bibinfo  {journal} {Phys. Rev. Lett.}\ }\textbf {\bibinfo {volume} {91}},\
  \bibinfo {pages} {196601} (\bibinfo {year} {2003})}\BibitemShut {NoStop}%
\bibitem [{\citenamefont {Beenakker}\ \emph {et~al.}(2003)\citenamefont
  {Beenakker}, \citenamefont {Kindermann},\ and\ \citenamefont
  {Nazarov}}]{Beenakker2003}%
  \BibitemOpen
  \bibfield  {author} {\bibinfo {author} {\bibfnamefont {C.~W.~J.}\
  \bibnamefont {Beenakker}}, \bibinfo {author} {\bibfnamefont {M.}~\bibnamefont
  {Kindermann}},\ and\ \bibinfo {author} {\bibfnamefont {Y.~V.}\ \bibnamefont
  {Nazarov}},\ }\bibfield  {title} {\bibinfo {title} {Temperature-dependent
  third cumulant of tunneling noise},\ }\href
  {https://doi.org/10.1103/PhysRevLett.90.176802} {\bibfield  {journal}
  {\bibinfo  {journal} {Phys. Rev. Lett.}\ }\textbf {\bibinfo {volume} {90}},\
  \bibinfo {pages} {176802} (\bibinfo {year} {2003})}\BibitemShut {NoStop}%
\bibitem [{\citenamefont {Gershon}\ \emph {et~al.}(2008)\citenamefont
  {Gershon}, \citenamefont {Bomze}, \citenamefont {Sukhorukov},\ and\
  \citenamefont {Reznikov}}]{Gershon2008}%
  \BibitemOpen
  \bibfield  {author} {\bibinfo {author} {\bibfnamefont {G.}~\bibnamefont
  {Gershon}}, \bibinfo {author} {\bibfnamefont {Y.}~\bibnamefont {Bomze}},
  \bibinfo {author} {\bibfnamefont {E.~V.}\ \bibnamefont {Sukhorukov}},\ and\
  \bibinfo {author} {\bibfnamefont {M.}~\bibnamefont {Reznikov}},\ }\bibfield
  {title} {\bibinfo {title} {Detection of non-gaussian fluctuations in a
  quantum point contact},\ }\href
  {https://doi.org/10.1103/PhysRevLett.101.016803} {\bibfield  {journal}
  {\bibinfo  {journal} {Phys. Rev. Lett.}\ }\textbf {\bibinfo {volume} {101}},\
  \bibinfo {pages} {016803} (\bibinfo {year} {2008})}\BibitemShut {NoStop}%
\bibitem [{\citenamefont {Nagaev}(2002)}]{Nagaev2002}%
  \BibitemOpen
  \bibfield  {author} {\bibinfo {author} {\bibfnamefont {K.~E.}\ \bibnamefont
  {Nagaev}},\ }\bibfield  {title} {\bibinfo {title} {Cascade boltzmann-langevin
  approach to higher-order current correlations in diffusive metal contacts},\
  }\href {https://doi.org/10.1103/PhysRevB.66.075334} {\bibfield  {journal}
  {\bibinfo  {journal} {Phys. Rev. B}\ }\textbf {\bibinfo {volume} {66}},\
  \bibinfo {pages} {075334} (\bibinfo {year} {2002})}\BibitemShut {NoStop}%
\bibitem [{\citenamefont {Gabelli}\ and\ \citenamefont
  {Reulet}(2008)}]{Gabelli2008}%
  \BibitemOpen
  \bibfield  {author} {\bibinfo {author} {\bibfnamefont {J.}~\bibnamefont
  {Gabelli}}\ and\ \bibinfo {author} {\bibfnamefont {B.}~\bibnamefont
  {Reulet}},\ }\bibfield  {title} {\bibinfo {title} {Dynamics of quantum noise
  in a tunnel junction under ac excitation},\ }\href
  {https://doi.org/10.1103/PhysRevLett.100.026601} {\bibfield  {journal}
  {\bibinfo  {journal} {Phys. Rev. Lett.}\ }\textbf {\bibinfo {volume} {100}},\
  \bibinfo {pages} {026601} (\bibinfo {year} {2008})}\BibitemShut {NoStop}%
\bibitem [{\citenamefont {Gabelli}\ and\ \citenamefont
  {Reulet}(2007)}]{Gabelli_SPIE}%
  \BibitemOpen
  \bibfield  {author} {\bibinfo {author} {\bibfnamefont {J.}~\bibnamefont
  {Gabelli}}\ and\ \bibinfo {author} {\bibfnamefont {B.}~\bibnamefont
  {Reulet}},\ }\bibfield  {title} {\bibinfo {title} {{The noise susceptibility
  of a coherent conductor}},\ }in\ \href {https://doi.org/10.1117/12.724656}
  {\emph {\bibinfo {booktitle} {Noise and Fluctuations in Circuits, Devices,
  and Materials}}},\ Vol.\ \bibinfo {volume} {6600},\ \bibinfo {organization}
  {International Society for Optics and Photonics}\ (\bibinfo  {publisher}
  {SPIE},\ \bibinfo {year} {2007})\ p.\ \bibinfo {pages} {66000T}\BibitemShut
  {NoStop}%
\bibitem [{\citenamefont {Gabelli}\ \emph {et~al.}(2013)\citenamefont
  {Gabelli}, \citenamefont {Spietz}, \citenamefont {Aumentado},\ and\
  \citenamefont {Reulet}}]{Gabelli2013}%
  \BibitemOpen
  \bibfield  {author} {\bibinfo {author} {\bibfnamefont {J.}~\bibnamefont
  {Gabelli}}, \bibinfo {author} {\bibfnamefont {L.}~\bibnamefont {Spietz}},
  \bibinfo {author} {\bibfnamefont {J.}~\bibnamefont {Aumentado}},\ and\
  \bibinfo {author} {\bibfnamefont {B.}~\bibnamefont {Reulet}},\ }\bibfield
  {title} {\bibinfo {title} {Electron–photon correlations and the third
  moment of quantum noise},\ }\href
  {https://doi.org/10.1088/1367-2630/15/11/113045} {\bibfield  {journal}
  {\bibinfo  {journal} {New Journal of Physics}\ }\textbf {\bibinfo {volume}
  {15}},\ \bibinfo {pages} {113045} (\bibinfo {year} {2013})}\BibitemShut
  {NoStop}%
\bibitem [{\citenamefont {Farley}\ \emph {et~al.}(2023)\citenamefont {Farley},
  \citenamefont {Pinsolle},\ and\ \citenamefont {Reulet}}]{Farley2023}%
  \BibitemOpen
  \bibfield  {author} {\bibinfo {author} {\bibfnamefont {C.}~\bibnamefont
  {Farley}}, \bibinfo {author} {\bibfnamefont {E.}~\bibnamefont {Pinsolle}},\
  and\ \bibinfo {author} {\bibfnamefont {B.}~\bibnamefont {Reulet}},\
  }\bibfield  {title} {\bibinfo {title} {Noise dynamics in the quantum
  regime},\ }in\ \href {https://doi.org/10.1109/ICNF57520.2023.10472766} {\emph
  {\bibinfo {booktitle} {2023 International Conference on Noise and
  Fluctuations (ICNF)}}}\ (\bibinfo {year} {2023})\ pp.\ \bibinfo {pages}
  {1--4}\BibitemShut {NoStop}%
\bibitem [{\citenamefont {Galaktionov}\ \emph {et~al.}(2003)\citenamefont
  {Galaktionov}, \citenamefont {Golubev},\ and\ \citenamefont
  {Zaikin}}]{Galaktionov2003}%
  \BibitemOpen
  \bibfield  {author} {\bibinfo {author} {\bibfnamefont {A.~V.}\ \bibnamefont
  {Galaktionov}}, \bibinfo {author} {\bibfnamefont {D.~S.}\ \bibnamefont
  {Golubev}},\ and\ \bibinfo {author} {\bibfnamefont {A.~D.}\ \bibnamefont
  {Zaikin}},\ }\bibfield  {title} {\bibinfo {title} {Statistics of current
  fluctuations in mesoscopic coherent conductors at nonzero frequencies},\
  }\href {https://doi.org/10.1103/PhysRevB.68.235333} {\bibfield  {journal}
  {\bibinfo  {journal} {Phys. Rev. B}\ }\textbf {\bibinfo {volume} {68}},\
  \bibinfo {pages} {235333} (\bibinfo {year} {2003})}\BibitemShut {NoStop}%
\bibitem [{\citenamefont {Salo}\ \emph {et~al.}(2006)\citenamefont {Salo},
  \citenamefont {Hekking},\ and\ \citenamefont {Pekola}}]{Salo2006}%
  \BibitemOpen
  \bibfield  {author} {\bibinfo {author} {\bibfnamefont {J.}~\bibnamefont
  {Salo}}, \bibinfo {author} {\bibfnamefont {F.~W.~J.}\ \bibnamefont
  {Hekking}},\ and\ \bibinfo {author} {\bibfnamefont {J.~P.}\ \bibnamefont
  {Pekola}},\ }\bibfield  {title} {\bibinfo {title} {Frequency-dependent
  current correlation functions from scattering theory},\ }\href
  {https://doi.org/10.1103/PhysRevB.74.125427} {\bibfield  {journal} {\bibinfo
  {journal} {Phys. Rev. B}\ }\textbf {\bibinfo {volume} {74}},\ \bibinfo
  {pages} {125427} (\bibinfo {year} {2006})}\BibitemShut {NoStop}%
\bibitem [{\citenamefont {Bednorz}\ and\ \citenamefont
  {Belzig}(2010{\natexlab{b}})}]{Bednorz2010}%
  \BibitemOpen
  \bibfield  {author} {\bibinfo {author} {\bibfnamefont {A.}~\bibnamefont
  {Bednorz}}\ and\ \bibinfo {author} {\bibfnamefont {W.}~\bibnamefont
  {Belzig}},\ }\bibfield  {title} {\bibinfo {title} {Models of mesoscopic
  time-resolved current detection},\ }\href
  {https://doi.org/10.1103/PhysRevB.81.125112} {\bibfield  {journal} {\bibinfo
  {journal} {Phys. Rev. B}\ }\textbf {\bibinfo {volume} {81}},\ \bibinfo
  {pages} {125112} (\bibinfo {year} {2010}{\natexlab{b}})}\BibitemShut
  {NoStop}%
\bibitem [{\citenamefont {Blanter}\ and\ \citenamefont
  {Büttiker}(2000)}]{Blanter2000}%
  \BibitemOpen
  \bibfield  {author} {\bibinfo {author} {\bibfnamefont {Y.}~\bibnamefont
  {Blanter}}\ and\ \bibinfo {author} {\bibfnamefont {M.}~\bibnamefont
  {Büttiker}},\ }\bibfield  {title} {\bibinfo {title} {Shot noise in
  mesoscopic conductors},\ }\href
  {https://doi.org/https://doi.org/10.1016/S0370-1573(99)00123-4} {\bibfield
  {journal} {\bibinfo  {journal} {Physics Reports}\ }\textbf {\bibinfo {volume}
  {336}},\ \bibinfo {pages} {1} (\bibinfo {year} {2000})}\BibitemShut {NoStop}%
\bibitem [{\citenamefont {Dolan}(1977)}]{Dolan1977}%
  \BibitemOpen
  \bibfield  {author} {\bibinfo {author} {\bibfnamefont {G.~J.}\ \bibnamefont
  {Dolan}},\ }\bibfield  {title} {\bibinfo {title} {Offset masks for lift‐off
  photoprocessing},\ }\href {https://doi.org/10.1063/1.89690} {\bibfield
  {journal} {\bibinfo  {journal} {Applied Physics Letters}\ }\textbf {\bibinfo
  {volume} {31}},\ \bibinfo {pages} {337} (\bibinfo {year} {1977})}\BibitemShut
  {NoStop}%
\bibitem [{\citenamefont {Spietz}\ \emph {et~al.}(2003)\citenamefont {Spietz},
  \citenamefont {Lehnert}, \citenamefont {Siddiqi},\ and\ \citenamefont
  {Schoelkopf}}]{Spietz2003}%
  \BibitemOpen
  \bibfield  {author} {\bibinfo {author} {\bibfnamefont {L.}~\bibnamefont
  {Spietz}}, \bibinfo {author} {\bibfnamefont {K.~W.}\ \bibnamefont {Lehnert}},
  \bibinfo {author} {\bibfnamefont {I.}~\bibnamefont {Siddiqi}},\ and\ \bibinfo
  {author} {\bibfnamefont {R.~J.}\ \bibnamefont {Schoelkopf}},\ }\bibfield
  {title} {\bibinfo {title} {Primary electronic thermometry using the shot
  noise of a tunnel junction},\ }\href
  {https://doi.org/10.1126/science.1084647} {\bibfield  {journal} {\bibinfo
  {journal} {Science}\ }\textbf {\bibinfo {volume} {300}},\ \bibinfo {pages}
  {1929} (\bibinfo {year} {2003})}\BibitemShut {NoStop}%
\bibitem [{\citenamefont {F\'evrier}\ \emph {et~al.}(2020)\citenamefont
  {F\'evrier}, \citenamefont {Lupien},\ and\ \citenamefont {Reulet}}]{S3cyclo}%
  \BibitemOpen
  \bibfield  {author} {\bibinfo {author} {\bibfnamefont {P.}~\bibnamefont
  {F\'evrier}}, \bibinfo {author} {\bibfnamefont {C.}~\bibnamefont {Lupien}},\
  and\ \bibinfo {author} {\bibfnamefont {B.}~\bibnamefont {Reulet}},\
  }\bibfield  {title} {\bibinfo {title} {Fundamental and environmental
  contributions to the cyclostationary third moment of current fluctuations in
  a tunnel junction},\ }\href {https://doi.org/10.1103/PhysRevB.101.245440}
  {\bibfield  {journal} {\bibinfo  {journal} {Phys. Rev. B}\ }\textbf {\bibinfo
  {volume} {101}},\ \bibinfo {pages} {245440} (\bibinfo {year}
  {2020})}\BibitemShut {NoStop}%
\bibitem [{\citenamefont {Esposito}\ \emph {et~al.}(2021)\citenamefont
  {Esposito}, \citenamefont {Ranadive}, \citenamefont {Planat},\ and\
  \citenamefont {Roch}}]{TWPAs}%
  \BibitemOpen
  \bibfield  {author} {\bibinfo {author} {\bibfnamefont {M.}~\bibnamefont
  {Esposito}}, \bibinfo {author} {\bibfnamefont {A.}~\bibnamefont {Ranadive}},
  \bibinfo {author} {\bibfnamefont {L.}~\bibnamefont {Planat}},\ and\ \bibinfo
  {author} {\bibfnamefont {N.}~\bibnamefont {Roch}},\ }\bibfield  {title}
  {\bibinfo {title} {{Perspective on traveling wave microwave parametric
  amplifiers}},\ }\href {https://doi.org/10.1063/5.0064892} {\bibfield
  {journal} {\bibinfo  {journal} {Applied Physics Letters}\ }\textbf {\bibinfo
  {volume} {119}},\ \bibinfo {pages} {120501} (\bibinfo {year}
  {2021})}\BibitemShut {NoStop}%
\bibitem [{\citenamefont {Spietz}\ \emph {et~al.}(2006)\citenamefont {Spietz},
  \citenamefont {Schoelkopf},\ and\ \citenamefont {Pari}}]{Spietz2006}%
  \BibitemOpen
  \bibfield  {author} {\bibinfo {author} {\bibfnamefont {L.}~\bibnamefont
  {Spietz}}, \bibinfo {author} {\bibfnamefont {R.~J.}\ \bibnamefont
  {Schoelkopf}},\ and\ \bibinfo {author} {\bibfnamefont {P.}~\bibnamefont
  {Pari}},\ }\bibfield  {title} {\bibinfo {title} {{Shot noise thermometry down
  to 10mK}},\ }\href {https://doi.org/10.1063/1.2382736} {\bibfield  {journal}
  {\bibinfo  {journal} {Applied Physics Letters}\ }\textbf {\bibinfo {volume}
  {89}},\ \bibinfo {pages} {183123} (\bibinfo {year} {2006})}\BibitemShut
  {NoStop}%
\bibitem [{\citenamefont {Souquet}\ \emph {et~al.}(2014)\citenamefont
  {Souquet}, \citenamefont {Woolley}, \citenamefont {Gabelli}, \citenamefont
  {Simon},\ and\ \citenamefont {Clerk}}]{Souquet2014}%
  \BibitemOpen
  \bibfield  {author} {\bibinfo {author} {\bibfnamefont {J.-R.}\ \bibnamefont
  {Souquet}}, \bibinfo {author} {\bibfnamefont {M.~J.}\ \bibnamefont
  {Woolley}}, \bibinfo {author} {\bibfnamefont {J.}~\bibnamefont {Gabelli}},
  \bibinfo {author} {\bibfnamefont {P.}~\bibnamefont {Simon}},\ and\ \bibinfo
  {author} {\bibfnamefont {A.~A.}\ \bibnamefont {Clerk}},\ }\bibfield  {title}
  {\bibinfo {title} {Photon-assisted tunnelling with nonclassical light},\
  }\href {https://doi.org/10.1038/ncomms6562} {\bibfield  {journal} {\bibinfo
  {journal} {Nature Communications}\ }\textbf {\bibinfo {volume} {5}},\
  \bibinfo {pages} {5562} (\bibinfo {year} {2014})}\BibitemShut {NoStop}%
\end{thebibliography}%

\end{document}